\documentclass[superscriptaddress,prx,twocolumn]{revtex4-1}
\usepackage{graphicx}
\usepackage{amsmath,amssymb,physics,dsfont}
\usepackage{color}
\usepackage{hyperref}
\usepackage{microtype}
\usepackage{braket}
\usepackage{defs-private}

\begin{document}

\title{Adaptive Variational Quantum Dynamics Simulations}
\author{Yong-Xin Yao}
\email{ykent@iastate.edu}
\affiliation{Ames Laboratory, U.S. Department of Energy, Ames, Iowa 50011, USA}
\affiliation{Department of Physics and Astronomy, Iowa State University, Ames, Iowa 50011, USA}

\author{Niladri Gomes}
\affiliation{Ames Laboratory, U.S. Department of Energy, Ames, Iowa 50011, USA}

\author{Feng Zhang}
\affiliation{Ames Laboratory, U.S. Department of Energy, Ames, Iowa 50011, USA}

\author{Cai-Zhuang Wang}
\affiliation{Ames Laboratory, U.S. Department of Energy, Ames, Iowa 50011, USA}
\affiliation{Department of Physics and Astronomy, Iowa State University, Ames, Iowa 50011, USA}

\author{Kai-Ming Ho}
\affiliation{Ames Laboratory, U.S. Department of Energy, Ames, Iowa 50011, USA}
\affiliation{Department of Physics and Astronomy, Iowa State University, Ames, Iowa 50011, USA}

\author{Thomas Iadecola}
\affiliation{Ames Laboratory, U.S. Department of Energy, Ames, Iowa 50011, USA}
\affiliation{Department of Physics and Astronomy, Iowa State University, Ames, Iowa 50011, USA}

\author{Peter P. Orth}
\affiliation{Ames Laboratory, U.S. Department of Energy, Ames, Iowa 50011, USA}
\affiliation{Department of Physics and Astronomy, Iowa State University, Ames, Iowa 50011, USA}

\begin{abstract}
	  We propose a general-purpose, self-adaptive approach to construct variational wavefunction ans\"atze for highly accurate quantum dynamics simulations based on McLachlan's variational principle.  The key idea is to dynamically expand the variational ansatz along the time-evolution path such that the ``McLachlan distance'', which is a measure of the simulation accuracy, remains below a set threshold. We apply this adaptive variational quantum dynamics simulation (AVQDS) approach to the integrable Lieb-Schultz-Mattis spin chain and the nonintegrable mixed-field Ising model, where it captures both finite-rate and sudden post-quench dynamics with high fidelity. The AVQDS quantum circuits that prepare the time-evolved state are much shallower than those obtained from first-order Trotterization and contain up to two orders of magnitude fewer CNOT gate operations. We envision that a wide range of dynamical simulations of quantum many-body systems on near-term quantum computing devices will be made possible through the AVQDS framework.
\end{abstract}

\maketitle

\section{Introduction}
One of the primary scientific quantum computing focuses has been to simulate ground-state, excited-states, and dynamical properties of spin and fermion systems~\cite{feynman82Simulating, lloyd96universal, Abrams97simulation, Abrams99Quantum, somma03quantum, Aspuru05Simulated, kassal08polynomial, Georgescu14Quantum, TroyerQCMB, Trotter_dynamics_Lawrence, Trotter_dynamics_Knolle, Cao19Quantum, McArdle20Quantum}. The ultimate goal is to accurately model physical systems of classically prohibitive size by efficient encoding and manipulation of many-body states. For ideal fault-tolerant quantum computers where deep circuits can be executed, algorithms built on Trotterized adiabatic state preparation and dynamics simulations, along with quantum phase estimation, can address a wide class of physical and chemical problems in and out of equilibrium~\cite{Abrams97simulation, Abrams99Quantum, kassal08polynomial, Trotter_dynamics_Lawrence, Trotter_dynamics_Knolle, McArdle20Quantum}. 

In the near term with noisy intermediate-scale quantum (NISQ) computers~\cite{nisq}, practical calculations on quantum devices are limited to short circuits. To exploit the emerging NISQ technology, the variational quantum eigensolver (VQE) has been developed and demonstrated to prepare the ground state of a time-independent Hamiltonian~\cite{vqe, vqe_theory, vqe_pea_h2, hardware_efficient_vqe, VQE_qcc, FengVQE}, or generally to minimize a static cost function with individual terms whose expectation values can be measured efficiently on quantum devices~\cite{qaoa}. Preconstructed fixed-form variational ans\"atze, such as the unitary coupled-cluster ansatz~\cite{vqe_theory, vqe_pea_h2, vqe_uccsd}, have been employed in early VQE calculations of molecules on quantum devices. However, the accuracy is often limited by the form of the ansatz~\cite{vqe_adaptive}, and the number of variational parameters and the circuit depth can grow as a high-order polynomial as the number of orbitals and atomic sites increases~\cite{vqe_theory, alan_ucc2018}. To alleviate these issues arising from a fixed form of the variational wavefunction, several \emph{adaptive} VQE methods have been proposed. These methods use a form of the variational wavefunction that is adaptively optimized for the specific problem~\cite{vqe_adaptive, vqe_qubit_adaptive}, leading to highly accurate results with much simpler variational circuits.

Variational approaches to quantum dynamics simulations (VQDS), including fast-forwarding methods, have also been proposed and applied to quantum spin models~\cite{theory_vqs, Endo20variational, cirstoiu2020variational, commeau2020variational, benedetti2020hardware}, with proof-of-principle applications on real devices~\cite{chen19demonstration}. The proposed variational wavefunction forms are described by a set of relatively shallow quantum circuits, which are tailored for execution on NISQ devices. 
The quality of variational quantum simulations is tied to the ability of the variational ansatz in describing the time-evolved wavefunction. For VQDS, the ansatz should be flexible enough to represent the quantum state along its time-evolution path. This is typically much more challenging than constructing an accurate variational ansatz for the ground state, because the nature of the wavefunction can change significantly during time evolution. Attempts to construct ans\"atze of fixed variational circuits for VQDS have been reported~\cite{chen19demonstration}, but the simulation accuracy quickly deteriorates as the system grows from $2$ sites to a few sites. This highlights the need for flexible variational circuits that can adapt to the changes of the wavefunction during time-evolution, while still keeping the circuit sufficiently shallow to be run on NISQ quantum processing units (QPUs). 

Here, we propose a novel time-dependent adaptive variational method to perform accurate quantum dynamics simulations of fermionic and quantum spin models. Going beyond a variational approach with a fixed cost function, the proposed scheme constructs an efficient time-dependent variational ansatz of the time-evolved wavefunction. This directly addresses the challenge of constructing efficient variational ans\"atze for dynamics simulations. This is generally a difficult task as the time-evolved wavefunction explores different, and a priori unknown, regions of Hilbert space. The proposed adaptive variational quantum dynamics simulation (AVQDS) approach is built on the McLachlan variational principle for real-time quantum dynamics simulations~\cite{mclachlan64variational, theory_vqs}, and automatically generates and dynamically expands the variational ansatz by minimizing the McLachlan distance $L^2$ along the time-evolution path. The form of the variational ansatz is incrementally expanded by choosing optimal operators from a predefined operator pool to construct additional unitary gates, in the same spirit as in the (static) ADAPT-VQE method~\cite{vqe_adaptive, vqe_qubit_adaptive}. The crucial difference is that here we optimize a time-dependent cost function, the McLachlan's distance $L^2(t)$ at time $t$. This allows the complexity of the ansatz to increase as needed during the time-evolution in order to accurately represent the wavefunction. It is worth noting that our key idea of adaptively generating variational ans\"atze by minimizing a \emph{time-dependent cost function} for dynamics simulations is widely applicable beyond the McLachlan approach. For example, it can also be used to generalize the projected-Variational Quantum Dynamics (p-VQD) simulation method discussed in Ref.~\cite{barison2021efficient} by replacing the McLachlan distance with the step-infidelity function.

We apply AVQDS to study linear ramp quantum dynamics of the integrable Lieb-Schultz-Mattis (LSM) spin model~\cite{lieb1961two, damle1996multicritical}, and sudden quench dynamics of the nonintegrable mixed-field Ising model (MFIM). In both cases, we find that dynamical quantities of interest (such as local observables, total energy and the Loschmidt echo) are described accurately with the adaptively generated variational ansatz. Notably, the state preparation circuits for the time-dependent wavefunction require two orders of magnitude fewer two-qubit gates than first-order Trotter dynamics simulations that achieve comparable accuracy. We demonstrate an initial time $t$-linear growth of the number of CNOT gates $N_\text{cx} \propto t$ in the AVQDS circuits, before $N_\text{cx}$ saturates after a critical time $t_\text{s}$. The saturation time $t_\text{s}$ is found to increase with system size $N$. The system size scaling of the \emph{saturated} $N_\text{cx}(t>t_\text{s})$ changes from quadratic ($\propto N^2$) to higher order ($\propto N^\alpha$ with $\alpha > 4$) as the integrability of the model is broken. In contrast, at fixed simulation times $t < t_\text{s}$, $N_\text{cx}$ scales approximately linearly with large $N > N_\text{c}(t)$ (where the circuits are in the presaturation regime). The crossover system-size $N_\text{c}$ grows slowly with $t$. This implies the practical scalability of general AVQDS simulations over finite time intervals.

The paper is organized as follows. The AVQDS algorithm, calculation procedures and important technical details are presented in Sec.~\ref{sec: method}. For completeness, the section also contains brief reviews of the Trotterized dynamics and the previously introduced VQDS formalism. Sec.~\ref{sec: results} discusses results for the integrable LSM model using a linear-ramp quench protocol and the observed gate count scaling of the AVQDS method. AVQDS simulations of quench dynamics in the nonintegrable MFIM are presented in Sec.~\ref{sec: mfim}. We summarize our results and give concluding remarks in Sec.~\ref{sec: conclusion}.

\begin{figure*}[t]
	\centering
	\includegraphics[width = \linewidth]{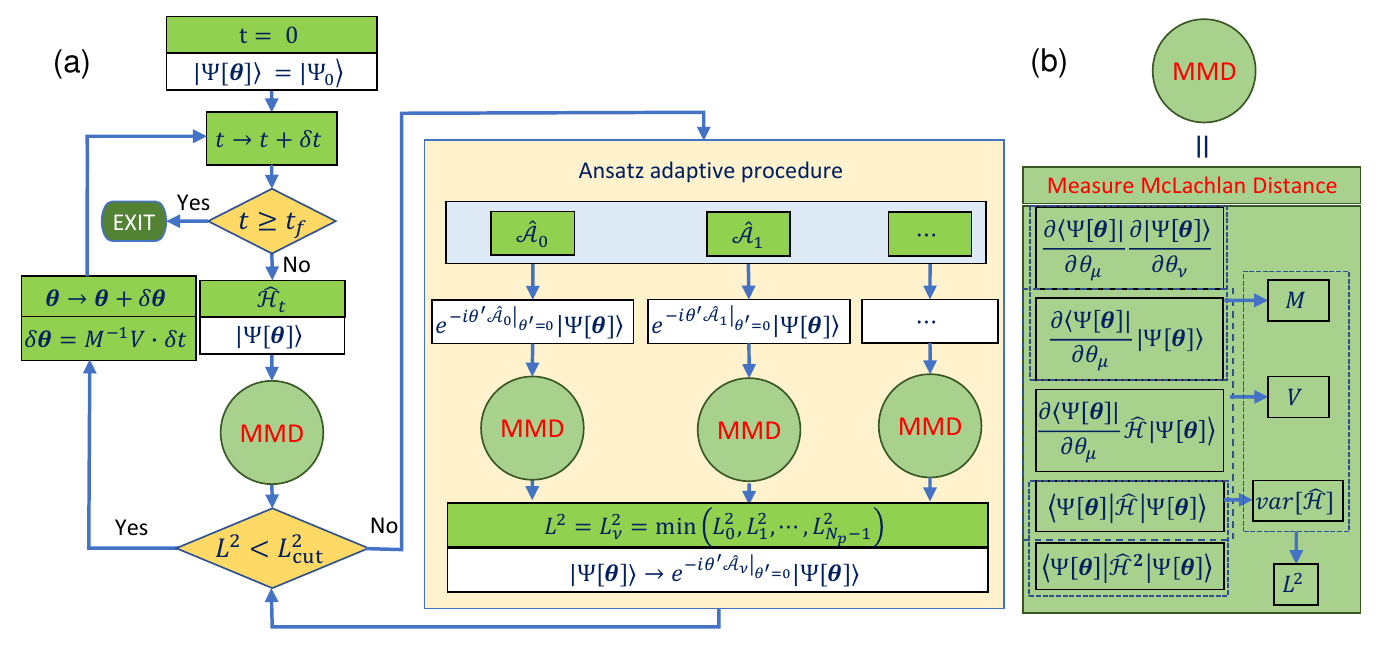}
	\caption{
	\textbf{Schematic illustration of the adaptive variational quantum dynamics simulation (AVQDS) algorithm.} The flow chart of AVQDS is plotted in panel (a). The details of the MMD module, which measures the McLachlan distance for a given variational wavefunction $\Psi[\bth]$ and time-dependent Hamiltonian $\h_t$, are shown in panel (b). The circuits to measure matrix $M$ and vector $V$ can be found in reference~\cite{theory_vqs} (see also Fig.~\ref{fig: imp}). Note that in the ansatz adaptive procedure, one only needs to measures the incremental elements in $M$ and $V$, which are added in a given step.
	}
	\label{algorithm}
\end{figure*}

\section{AVQDS method}
\label{sec: method}
\subsection{Trotterized state evolution}
For the convenience of later comparisons, we summarize the quantum dynamics simulation method using a discrete-time propagator based on the Trotter decomposition~\cite{trotter, nielsen2002quantum}. For a system described by a generic time-dependent Hamiltonian 
\be
\h[t] = \sum_{\mu}\hat{h}_\mu[t],
\ee
the quantum state evolves by a fixed time step $\delta t$ as
\be
\ket{\Psi[t+\delta t]} = \prod_{\mu}e^{-i\delta t \, \hat{h}_\mu[t]}\ket{\Psi[t]}. \label{eq: trotter}
\ee
Here, $e^{-i\delta t \, \hat{h}_\mu[t]}$ is a unitary that can be efficiently implemented on QPUs. Therefore, the Trotter circuit depth grows linearly with the number of time steps, with the incremental depth determined by the Hamiltonian terms $\{\hat{h}_\mu \}$. For calculations on quantum computers, each individual term $\hat{h}_\mu$ is a tensor product of Pauli operators. The exponential of a Pauli term coupling $p$ qubits scaled by an imaginary number constitutes a general Pauli rotation gate, which can be compiled to a series of one-qubit rotations and $2(p-1)$ two-qubit entangling gates for real quantum device applications~\cite{nielsen2002quantum}. 

Although it is not feasible to directly compile a general discrete-time Hamiltonian evolution operator in real quantum devices, on classical devices it can be more efficient to calculate the discrete-time evolution according to $\ket{\Psi[t+\delta t]} = e^{-i\delta t \, \h[t]}\ket{\Psi[t]}$. This approach is more tolerant to finite step sizes, and becomes exact for simulations of sudden quench dynamics where the system is evolved by a Hamiltonian that is piecewise constant in time. We adopt this approach to get numerically exact data for comparison with the simulation results to be discussed later.

\subsection{Variational Quantum Dynamics Simulation}
The formalism of variational quantum dynamics simulations (VQDS) has been systematically presented in Ref.~\cite{theory_vqs}. To facilitate later discussions, we summarize the main points of variational real-time dynamics simulations based on McLachlan's variational principle~\cite{mclachlan64variational}. We work at the level of the density matrix $\rho$, which eliminates the necessity of keeping track of the global phase of the state $\ket{\Psi}$. For a system in a pure state evolving under a time-dependent Hamiltonian $\h$, the density matrix $\rho=\ket{\Psi}\bra{\Psi}$ evolves according to the von-Neumann equation
\be
\frac{d\rho}{d t} = \Lag[\rho],
\ee
with $\Lag[\rho] = -i[\h, \rho]$. In the variational quantum simulation approach, the state $\ket{\Psi[\bth]}$ is parameterized by a real time-dependent variational parameter vector $\bth[t]$. Unlike variational quantum algorithms which optimise a high-dimensional static cost function through parameter learning, the variational parameters $\bth[t]$ will be updated according to the equations of motion derived by the McLachlan variational principle. Because all the observables and the related equations are implicitly time-dependent in dynamics simulations, we drop the parameter $t$ associated with the variables for simplicity. The McLachlan variational principle amounts to minimizing the distance $L$, or equivalently the squared distance, between the variationally evolving state and the exact propagating state, which is defined as
\begin{widetext}
\bea
L^2 &\equiv&\norm{\sum_\mu \frac{\partial \rho[\bth]}{\partial \theta_\mu} \dot{\theta}_\mu - \Lag[\rho]}^2_F
=\sum_{\mu \nu} M_{\mu \nu} \dot{\theta}_\mu \dot{\theta}_\nu -2 \sum_\mu V_\mu \dot{\theta}_\mu + \Tr[\Lag[\rho]^2]. \label{L2}
\eea
Here $\norm{\rho}_F \equiv \sqrt{\Tr[\rho^\dag \rho]}$ is the Frobenius norm of the matrix $\rho$. The matrix $M$ is real symmetric and defined as
\bea
M_{\mu \nu} &\equiv& \Tr[\frac{\partial \rho[\bth]}{\partial \theta_\mu}\frac{\partial \rho[\bth]}{\partial \theta_\nu}] = 2 \Re\left[\frac{\partial \bra{\Psi[\bth]}}{\partial \theta_\mu} \frac{\partial \ket{\Psi[\bth]}}{\partial \theta_\nu}\right. \notag + \left.\frac{\partial \bra{\Psi[\bth]}}{\partial \theta_\mu}\ket{\Psi[\bth]} \frac{\partial \bra{\Psi[\bth]}}{\partial \theta_\nu}\ket{\Psi[\bth]}\right], \label{eq: M}
\eea
which is equivalent to the quantum Fisher information matrix associated with the state fidelity~\cite{meyer2021fisher}.
The vector $V$ is given by
\bea
V_\mu &\equiv& \Tr[\Re\left[\frac{\partial \rho[\bth]}{\partial \theta_\mu} \Lag[\rho]\right]] = 2\Im\left[\frac{\partial \bra{\Psi[\bth]}}{\partial \theta_\mu}\h\ket{\Psi[\bth]} + \bra{\Psi[\bth]}\frac{\partial \ket{\Psi[\bth]}}{\partial \theta_\mu}\av{\h}_{\bth} \right], \label{eq: V}
\eea
where $\av{\h}_{\bth}\equiv \Av{\Psi[\bth]}{\h}$, and
\bea
\Tr[\Lag[\rho]^2] &=& 2\left(\av{\h^2}_{\bth} - \av{\h}_{\bth}^2 \right) = 2 \,  \text{var}_{\bth}[\h], \label{eq: var}
\eea
\end{widetext}
which describes the energy variance of $\h$ in the variational state $\ket{\Psi[\bth]}$. The second term in the bracket of Eqs.~\eqref{eq: M} and~\eqref{eq: V} originates from the global phase contribution~\cite{theory_vqs}.
In the geometric picture of quantum evolution~\cite{Anandan90geometry}, the integral of energy variance with respect to time is independent of the specific Hamiltonian and defines a distance along evolution path measured by the Fubini-Study metric~\cite{kobayashi63foundations}.

The minimization of the cost function Eq.~\eqref{L2} with respect to $\{\dot{\theta}_\mu \}$ leads to the following equation of motion for the variational parameters:
\be
\sum_\nu M_{\mu \nu}\dot{\theta}_\nu = V_\mu. \label{eq: eom}
\ee
The McLachlan distance $L^2$ of the variational ansatz $\Psi[\bth]$ at optimal parameter values can then be calculated as
\be
L^2 = 2\, \text{var}_{\bth}[\h]-\sum_{\mu \nu}V_\mu M^{-1}_{\mu \nu} V_\nu. \label{eq: L2min}
\ee
As pointed out in Ref.~\cite{theory_vqs}, $L^2$ provides a natural measure for the accuracy of quantum dynamics simulations.

\subsection{Adaptive Variational Quantum Dynamics Simulation (AVQDS) approach}
\subsubsection{Algorithm and flow chart}
The adaptive variational quantum dynamics simulation method is illustrated in Fig.~\ref{algorithm}. The technique dynamically constructs a variational ansatz of the following pseudo-Trotter form:
\be
\ket{\Psi[\bth]} = \prod_{\mu=0}^{N_{\bth} - 1} e^{-i\theta_\mu \hat{\A}_\mu} \ket{\Psi_0}, \label{eq: ansatz}
\ee
where $\ket{\Psi_0}$ is a reference state and $\{\theta_\mu\}$ ($\mu=0,\dots,N_{\bth}-1$) are the time-dependent variational parameters. $\hat{\A_\mu}$ is a Hermitian operator. The set of $N_{\bth}$ operators $\{\hat{\A}_\mu \}$ will be dynamically expanded by including additional operators from an operator pool to maintain the McLachlan distance $L^2$~\eqref{eq: L2min} below a threshold $L^2_{\text{cut}}$, if necessary, as the system evolves. In practice, we find $L^2_{\text{cut}}=10^{-3}$ is enough to get highly accurate results.

Without loss of generality, let the dynamics simulation start at $t=0$ with the system in a pure state $\rho_0 = \ket{\Psi_0}\bra{\Psi_0}$, which we choose as the reference state of the variational ansatz. More specifically, at $t=0$ the variational ansatz $\ket{\Psi[\bth]}=\ket{\Psi_0}$ has no variational parameters (i.e., $N_{\bth}=0$). After an incremental time step $t=\delta t$, the Hamiltonian becomes $\h_{t}$. The McLachlan distance $L^2$ is measured with the previously obtained ansatz state $\ket{\Psi[\bth]}$ through the MMD module, which is specified in Fig.~\ref{algorithm}(b) according to Eqs.~\eqref{eq: M}-~\eqref{eq: var}. In the initial case where the variational ansatz has no parameters, the McLachlan distance is determined by the energy variance of $\h_t$ only. The energy variance is generally larger than zero, because the ansatz state is not an eigenstate of $\h_t$ as the system evolves in time. The McLachlan distance $L^2$ is then compared against the threshold $L^2_{\text{cut}}$, and the ansatz adaptive procedure is triggered if $L^2 \geq L^2_{\text{cut}}$. 

The adaptive procedure starts with evaluating the McLachlan distance $L^2$ with respect to a series of new variational ans\"atze. Each new ansatz, $e^{-i\theta'\hat{\A}_\mu |_{\theta'=0}}\ket{\Psi[\bth]}$, is composed of the existing ansatz, multiplied by the exponential of an generator $\hat{\A}_\mu$ with a coefficient $\theta'$ initialized to zero at the current time step. Although the new ansatz with $\theta' = 0$ does not change the ansatz state, the McLachlan distance $L^2$ can change due to nonvanishing derivatives with respect to the additional parameter in Eqs.~\eqref{eq: M} and ~\eqref{eq: V}.  Here, the choice of $\hat{\A}_\mu$ runs through all the operators in a preconstructed (fixed) operator pool of size $N_p$. 

For each operator $e^{-i\theta' \hat{\A}_\mu}$ that is added to the ansatz, the dimension of $\bth$ increases from $N_{\bth}$ to $N_{\bth} + 1$. Accordingly, the dimension of the symmetric matrix $M$~\eqref{eq: M} increases from $N_{\bth} \times N_{\bth}$ to $(N_{\bth}+1) \times (N_{\bth}+1)$, and the dimension of the vector $V$~\eqref{eq: V} increases from $N_{\bth}$ to $N_{\bth}+1$. Because the additional parameter in the new ansatz is always initialized to zero, the represented ansatz state remains the same. Therefore, only the additional $(N_{\bth} + 1)^\text{th}$ row of the matrix $M$ and the final element of $V$ need to be evaluated. The obtained $\{L^2_\mu\}$  ($\mu=0, \dots, N_p - 1$) values are compared, and one selects the operator $\hat{\A}_\nu$ for which $L^2_\nu$ is minimal. The ansatz form is updated to $\ket{\Psi[\bth]} \rightarrow e^{-i\theta_\nu \hat{\A}_\nu}\ket{\Psi[\bth]}$, with $\theta_\nu$ initially set to zero and the number of variational parameters $N_{\bth}$ increased by one. Note that setting $\theta_{N_{\bth}+1} = 0$ initially ensures that the wavefunction is smooth during time evolution. The McLachlan distance $L^2$ is then updated and compared with the threshold $L^2_\text{cut}$. For $L^2 \geq L^2_\text{cut}$ the ansatz adaptive procedure is repeated until $L^2 < L^2_\text{cut}$ is satisfied. Only then the variational parameters are updated, $\bth \to \bth + \delta \bth$, at current time step according to 
\bea
\delta \bth &=& \dot{\bth} \delta t = M^{-1}V\, \delta t. \label{eq: theta}
\eea
In typical AVQDS simulations reported here, the number of unitaries added to the ansatz at some initial time steps can be as much as about $10$ per step to gain sufficient expressibility, but quickly decreases to about $1$ per step if the McLachlan distance $L^2$ goes beyond the threshold $L^2_\text{cut}$. Finally, the system evolves to the next time step and the algorithmic procedure continues until the simulation time period ends.

\subsubsection{Initial state preparation}
The AVQDS calculation starts with some initial state, which may often be the ground state $\ket{\Psi_0}$ of a Hamiltonian $\h_0$ at $t=0$. A general efficient state preparation algorithm for near-term quantum devices remains an open challenge and has attracted much research effort, including adiabatic state preparation~\cite{Aspuru05Simulated}, VQE~\cite{vqe_theory}, and quantum imaginary time evolution~\cite{qite_chan20, VQITE, smqite}. Naturally, one can replace the original proposal of Trotter dynamics-based adiabatic state preparation with the AVQDS approach. This corresponds to first performing an AVQDS simulation starting from a Hamiltonian with a tensor-product ground state and evolving to the Hamiltonian $\h_0$ with ground state $\ket{\Psi_0}$ at a sufficiently slow quench rate. Alternatively, the ground state of the initial Hamiltonian $\h_0$ can be obtained using (static) ADAPT-VQE~\cite{vqe_adaptive, vqe_qubit_adaptive}, which can easily be combined with AVQDS. More specifically, we use the recently proposed qubit-ADAPT-VQE technique~\cite{vqe_qubit_adaptive} to prepare the ground state $\Psi_0$ of the initial Hamiltonian $\h_0$. Compared with AVQDS, which dynamically updates the variational ansatz~\eqref{eq: ansatz} with the goal of minimizing McLachlan's distance along a time-evolution path, qubit-ADAPT-VQE uses an adaptive scheme to optimize the variational ansatz in Eq.~\eqref{eq: ansatz} to minimize the expectation value of a time-independent Hamiltonian. Note that when encoding the initial state in this way, the initial number of variational parameters in AVQDS is larger than zero. The dynamical ansatz adaptive procedure can of course be carried out in the same way as described above. \label{adapt-vqe}

In the AVQDS simulations reported below, we focus on the dynamics simulation part, where we dynamically construct the variational ansatz and update the parameters according to Eq.~\eqref{eq: theta}. Further details on state preparations using qubit-ADAPT-VQE and its combination with AVQDS will be given in Section~\ref{adapt-vqe-avqds}.

\begin{figure*}[t]
	\centering
	\includegraphics[width=0.98\linewidth]{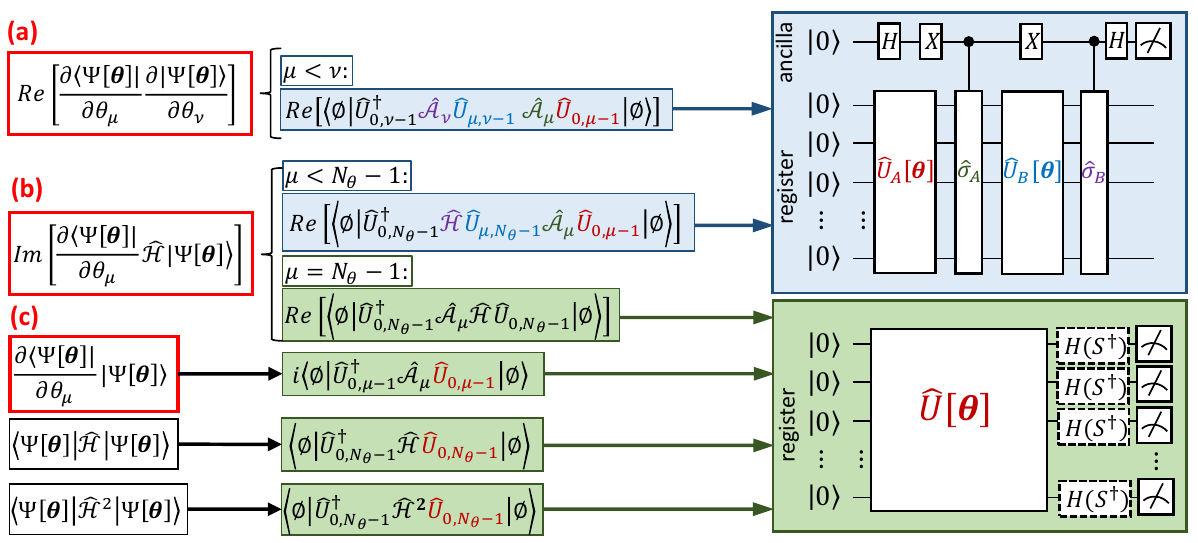}
	\caption{
	\textbf{Quantum circuit implementation of the AVQDS algorithm.} The left column lists the unique terms to be evaluated in Eqs.~\eqref{eq: M}-~\eqref{eq: var} of VQDS, with the terms (a, b, c) highlighted in red also involved in the ansatz adaptive procedure in AVQDS. The middle column specifies the expressions when the wavefunction ansatz takes the pseudo-Trotter form of Eq.~\eqref{eq: ansatz} with $\hat{U}_{j,k}[\bth]=\prod_{\mu=j}^{k}e^{-i\theta_{\mu}\hat{\A}_{\mu}}$ and $\ket{\Psi_0}=\ket{\emptyset}\equiv \otimes_{j=0}^{N - 1} \ket{0}$ for an $N$-qubit system. Two types of quantum circuits are adopted: green block for the direct measurement circuit, and blue block for generalized Hadamard test circuit~\cite{theory_vqs, VDynamics_Li}. The direct measurement circuit includes optional Hadamard gate $H$ or Hadamard-Phase gate $HS^\dag$ when measuring $X$ or $Y$-Pauli strings present in  $\hat{\A}_{\mu}$, $\hat{\A}_{\mu}^2$, $\h$ and $\h^2$. Accordingly, $\hat{U}[\bth]$ can be $\hat{U}_{0, \mu-1}$ or $\hat{U}_{0, N_{\bth}-1}$ as highlighted in middle column. In the generalized Hadamard test, the expectation value $Re\left[ \Av{\emptyset}{\hat{U}_A^\dag[\bth]\hat{U}_B^\dag[\bth] \hat{\sigma}_B\hat{U}_B[\bth]\hat{\sigma}_A\hat{U}_A[\bth]} \right]$ is given by $2P_{\ket{0}} -1 $, with $P_{\ket{0}}$ the probability that the ancillary qubit is in state $\ket{0}$. $\hat{U}_A[\bth]$ can be $\hat{U}_{0, \mu-1}$ or $\hat{U}_{0, \nu-1}[\bth]$ as dark red color encoded in the blue box of middle column, with similar color encoding for the other unitaries. $\hat{\sigma}_B$ represents a Pauli string in $\hat{\A}_{\mu}$ or $\h$, which is typically defined on one or two qubits for spin models. Therefore, the controlled-$\sigma$ gate represents a controlled-single qubit or controlled two-qubit gate. Alternatively, the generalized Hadamard test circuit can be replaced with direct measurement circuit~\cite{mitarai2019methodology, mari2021estimating}.
	}
	\label{fig: imp}
\end{figure*}

\subsubsection{Important technical details}
\label{dtheta_max}
Let us discuss some important technical details in the practical implementation of the AVQDS algorithm presented above, which can be used for dynamics simulations of generic Hamiltonian systems, including fermionic and spin systems. Since the system Hamiltonian will always be transformed to a qubit representation for calculations on QPUs, the operator pool $\{\hat{\A}_\mu \}$ can be constructed from a set of Pauli terms, i.e., tensor products of Pauli operators at different spin-orbital sites. The pool can be made of a complete list of Pauli terms up to some fixed length, or only involve Pauli terms that appear in the system Hamiltonian $\h$. Symmetries leading to conservation of particle number and spin quantum numbers, time-reversal symmetry and point group symmetries, can be considered to further reduce the pool size~\cite{bravyi2017tapering, setia2019reducing}. For fermionic systems, the operator pool can also be constructed using fermion operators, i.e., tensor products of fermion creation and annihilation operators, subject to symmetry constraints, before translating into qubit operators. This can potentially reduce the size of the operator pool for simulations of fermionic systems, at the cost that each operator maps, in general, to a sum of several Pauli terms. In the numerical calculations of spin models that we present below, we adopt the Hamiltonian pool which is composed of Pauli terms present in the Hamiltonian.

The AVQDS approach amounts to numerically integrating a system of ordinary differential equations~\eqref{eq: eom}. Within the well-known Euler method, the local truncation error at a single time step is of order of $(\delta t)^2$~\cite{iserles2009ode}, where $\delta t$ is the step size. This leads to a global truncation error over the total simulation period that scales linearly with $\delta t$. Higher-order methods such as the Runge–Kutta technique yield more favorable scaling~\cite{iserles2009ode}. In the numerical simulations presented here, we adopt the Euler method for simplicity and a fair comparison with the first order Trotter dynamics simulations.

It is useful to contrast the change of the wavefunction that occurs during one timestep within Trotterized and AVQDS simulations. Within Trotter dynamics, the size of the timestep $\delta t$ controls the change of the wavefunction during a single step (see Eq.~\eqref{eq: trotter}). In contrast, in AVQDS the change is controlled by the change of the variational parameters, $\delta \bth$ (see Eq.~\eqref{eq: theta}). This leads to an effective way to stabilize AVQDS simulations by fixing a maximal step size $\delta \theta_\text{max}$. In other words, the time step size $\delta t$ is dynamically adjusted at each time step such that $\text{max}_\mu (|\delta \theta_\mu|) \leq \delta \theta_\text{max}$. The dynamical time step size $\delta t$ in AVQDS is therefore set by $\delta \theta_\text{max}$ and the maximal absolute value $|\dot{\theta}|_\text{max}$ of elements in the vector $\dot{\bth}$ in Eq.~\eqref{eq: theta}. When comparing AVQDS to Trotterized dynamics simulations, $\delta \theta_\text{max}$ should be chosen to be the same as the fixed time step size $\delta t$ in Trotter dynamics to achieve a similar accuracy. An additional consequence is that $|\dot{\theta}|_\text{max}$ determines the total number of time steps required in AVQDS calculations to reach a given final time. 

We note that in the numerical simulations presented below, we find that the evaluation of $\dot{\bth} = M^{-1}V$ at intermediate time steps can involve a matrix $M$ with large condition number~\cite{suli2003introduction}, which are often encountered in numerical statistical analysis and machine learning~\cite{golden1996mathematical}. The issue can be alleviated by adding a small diagonal element ($\xi = 10^{-6}$) to the matrix $M$ before inversion following Tikhonov regularization, which effectively penalizes the large magnitude of $\dot{\bth}$. As a result, we find that $|\dot{\theta}|_\text{max}$ falls in the range of $(1, 10)$ in the following calculations. For AVQDS calculations in the presence of noise (both quantum mechanical shot noise and gate noise), we expect that the regularization parameter $\xi$ must be increased in order to deal with the resulting fluctuations in $M$ and $V$~\cite{qite_chan20, van2020measurement}. Alternatively, the matrix inversion in Eq.~\eqref{eq: theta} can be avoided altogether by solving the equation of motion~\eqref{eq: eom} using optimization techniques, which may also provide additional channels for gate error mitigation~\cite{endo2020calculation}. 

AVQDS simulations reported here are carried out with a classical implementation based on Quantum Toolbox in Python (QuTiP)~\cite{johansson2012qutip}. Because AVQDS relies on the same set of measurements as regular VQDS with a fixed wavefunction ansatz, the quantum circuit implementation on NISQ QPUs follows that of VQDS discussed previously~\cite{theory_vqs, VDynamics_Li}. Specifically, Fig.~\ref{fig: imp} lists the unique terms in determining the symmetric matrix $M$ in Eq.~\eqref{eq: M}, vector $V$ in Eq.~\eqref{eq: V} and scalars $\av{\h}_{\bth}$, $\av{\h^2}_{\bth}$ listed in the left column. The reduced expressions with the choice of pseudo-Trotter form~\eqref{eq: trotter} for the variational ansatz are listed in the middle column, which can be measured by direct measurement circuits (green) and generalized Hadamard test circuits (blue), as shown in the right column.

For an estimation of quantum resource of VQDS calculations, let us consider a Hamiltonian composed of $N_\text{H}$ Pauli strings and $N_{\bth}$ variational parameters in the wavefunction ansatz~\eqref{eq: ansatz}. The upper bound of the number of distinct direct measurement circuits and generalized Hadamard test circuits is $(N_\text{H}+2)N_{\bth} +N_\text{H} +N_\text{H}^2$ and $N_\text{H}(N_{\bth}-1) +(N_{\bth})(N_{\bth}-1)/2$, respectively. The ansatz adaptive expansion procedure in AVQDS adds a marginal overhead of quantum resources. The additional terms (a, b, c) as highlighted in Fig.~\ref{fig: imp} are to be measured only for unitaries to be appended in the variational circuit at the same state of the current time step, as discussed previously. Specifically, with a Hamiltonian operator pool composed of Hamiltonian terms as adopted in the following calculations, no additional measurements are required for terms (b, c) and part of terms (a), because all the contributions have already been measured when evaluating the expectation values of $\h$ and $\h^2$. Therefore the additional measurements in AVQDS from the ansatz adaptive procedure are for terms (a), $\Re\left[\frac{\partial \bra{\Psi[\bth]}}{\partial \theta_\mu} \frac{\partial \ket{\Psi[\bth]}}{\partial \theta_\nu}\right]$, with $\nu$ running through the Hamiltonian operator pool and $0\le \mu < N_{\bth}-1$, which amount to $N_\text{H}(N_{\bth}-1)$ generalized Hadamard test circuits. Each of the generalized Hadamard test circuits includes at most two controlled two-qubit gates for spin models presented here, which can also be replaced with direct measurement circuits~\cite{mitarai2019methodology, mari2021estimating}. For a system described by $N$ qubits with $N_\text{H}\propto N^p$, such as local spin models with $N_\text{H}\propto N$, and $N_{\bth} \propto N^q$, the leading order of distinct measurement circuits scales as $N^{2\max(p, q)}$, with the circuit depth tied to $N_{\bth}$. The overall measurement cost of AVQDS approach due to finite sampling shot noise follows the analysis of general metric-aware variational quantum algorithms~\cite{van2020measurement}. To keep the shot noise induced error below $\epsilon$, the number of samples generally scales as $\bigO (1/\epsilon^2)$, where the prefactor can be reduced by optimal measurement distribution~\cite{van2020measurement, Crawford_2021}.

\begin{figure}[ht]
	\centering
	\includegraphics[width=\columnwidth]{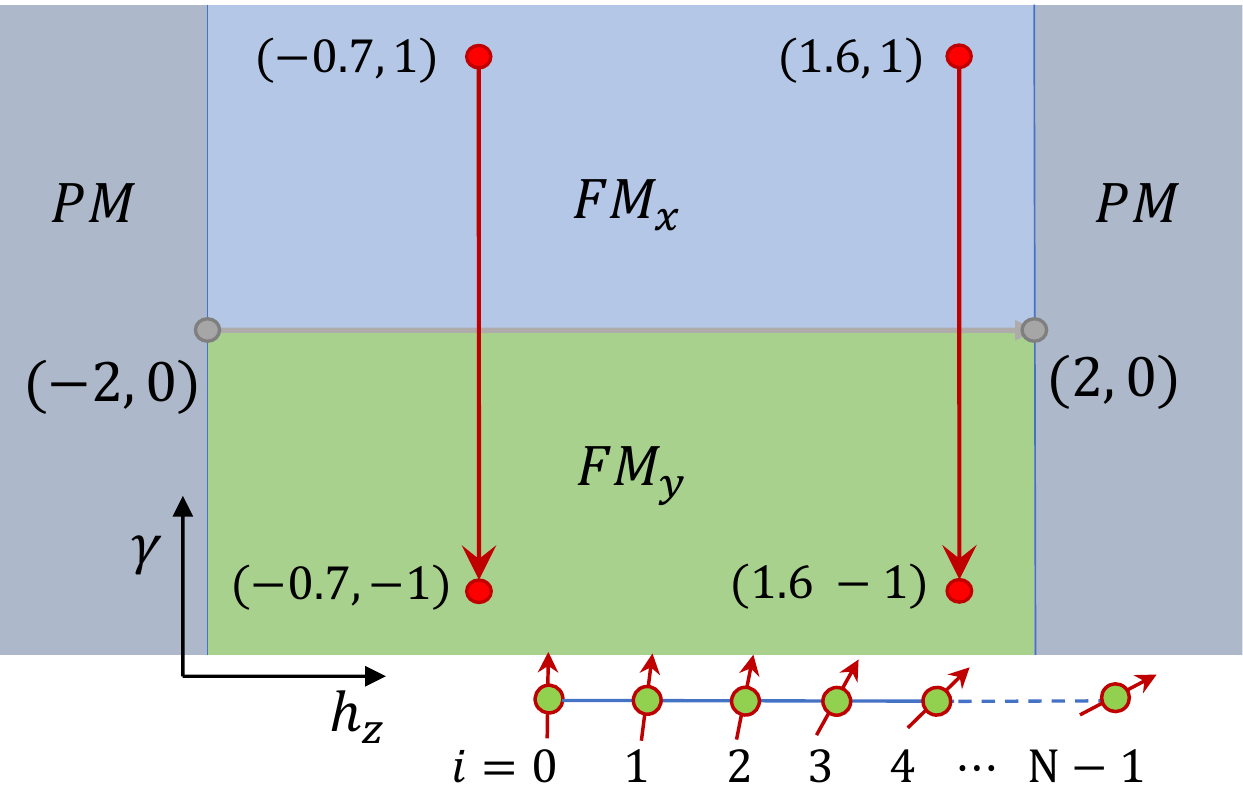}
	\caption{
	\textbf{Phase diagram of LSM model and the quench protocol.} Two ferromagnetic phases (FM$_x$, FM$_y$) and a paramagnetic phase are present in the ground state phase diagram of the LSM model in the thermodynamic limit ($N\rightarrow \infty$). The phase boundaries and tricritical points are also shown. Below we present results of AVQDS simulations for the two vertical parameter paths indicated here, which quenches the system from the FM$_x$ to the FM$_y$ phase at finite speed, together with post-quench dynamics.
	}
	\label{PhaseDiagram}
\end{figure}

\section{Linear ramp quantum dynamics in the LSM model}
\label{sec: results}
\subsection{Model and phase diagram}

\begin{figure*}[t]
	\centering
	\includegraphics[width=\textwidth]{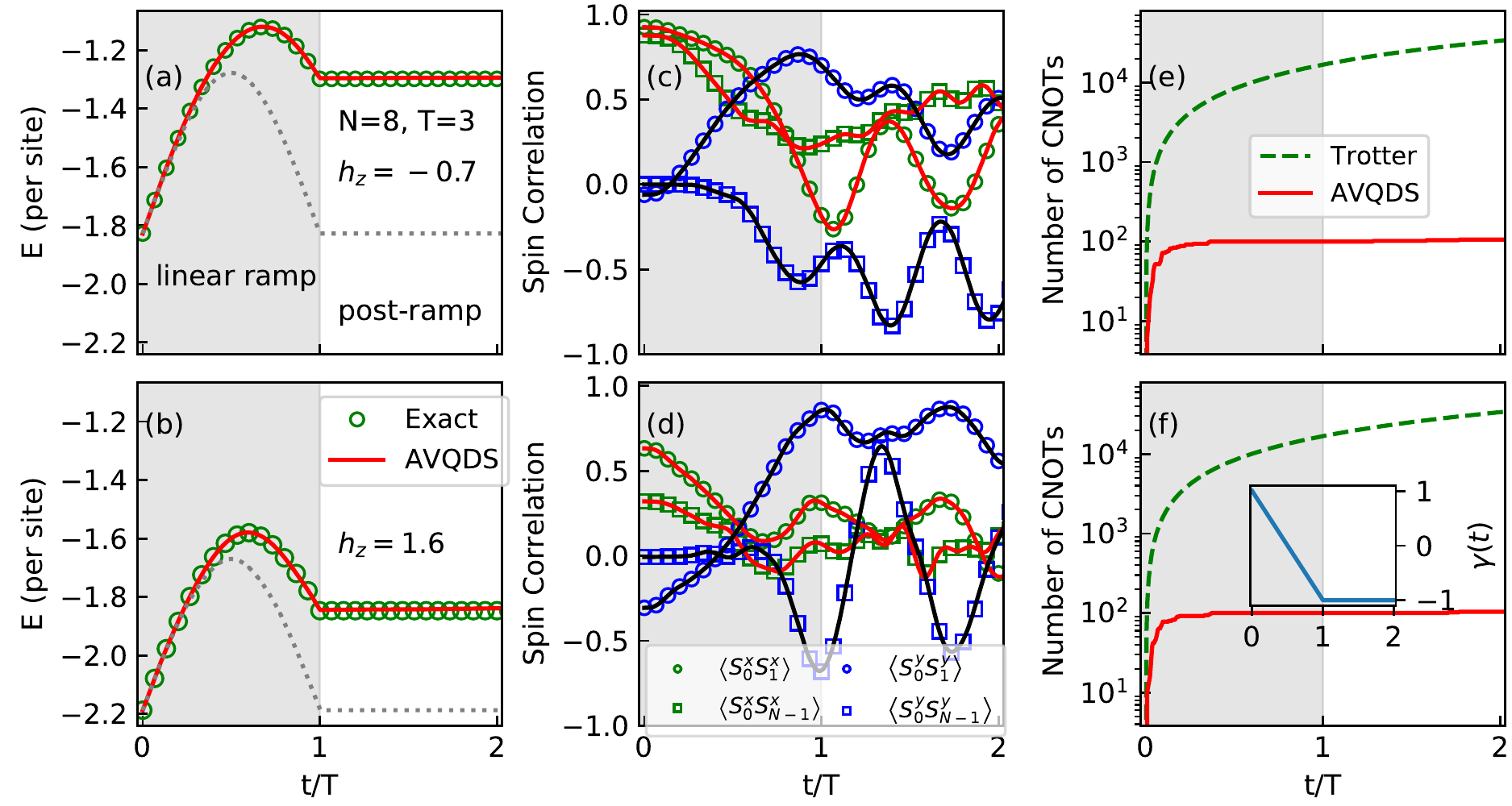}
	\caption{
	\textbf{Variations of instantaneous total energy, spin correlations and number of two-qubit gates for quantum quench simulations of the eight-site LSM chain model with open boundary conditions.} The simulation is composed of a linear ramp with speed defined by $T=3$, and post-ramp dynamics for another period of $T$. The results for $h_z = -0.7$ sites are shown in the upper panels, and that for $h_z = 1.6$ in the lower panels. The exact results are shown in symbols, and AVQDS data in solid curves. The evolution of the instantaneous energy from numerically exact calculations and AVQDS is shown in panels (a) and (b), along with the adiabatic results plotted as gray dotted lines for reference. The spin correlation function in panels (c) and (d) includes $x x$ and $y y$-components for a pair of the first and second site and a pair of the first and last site. Compared with linear circuit growth of Trotter dynamics simulations, AVQDS circuits are much shallower, with a sublinear circuit growth rate that quickly decreases, as plotted in panels (e) and (f). The ansatz at $t=T$ is composed of 50 two-qubit Pauli rotation gates (100 CNOTs) for both cases, which slightly increases to 53 two-qubit gates at $t=2T$. A multiplying factor of two for CNOT gates is included, since each two-qubit Pauli rotation gate can be compiled to two CNOTs along with single-qubit gates assuming full connectivity~\cite{nielsen2002quantum}. Inset in (f): variation of Hamiltonian parameter $\gamma$ as a function of $t$.
	}
	\label{escgates}
\end{figure*}

To demonstrate and benchmark the performance of the AVQDS method, we perform a series of finite-rate quantum quench dynamics simulations of the integrable $N$-site spin-$\frac{1}{2}$ Lieb-Schultz-Mattis chain with open boundary conditions, which describes an anisotropic XY Hamiltonian in a transverse magnetic field~\cite{lieb1961two}
\be
\h = - J \sum_{i=0}^{N-2} \left[ ( 1 + \gamma ) \hat{X}_i \hat{X}_{i+1} + (1 - \gamma) \hat{Y}_i \hat{Y}_{i+1} \right] + h_z \sum_{i=0}^{N-1} \hat{Z}_i,
\label{eq: XYZ}
\ee
where $\hat{X}, \hat{Y}$ and $\hat{Z}$ are single-site Pauli operators. The coupling constant $J$ is set to one as the energy unit. $h_z$ is the magnetic field strength. The anisotropy in the $xy$ plane is controlled by the parameter $\gamma$, and rotational symmetry around the z-axis is obtained with $\gamma=0$. The ground state phase diagram of the model in the thermodynamic limit ($N\rightarrow \infty$) is well known~\cite{damle1996multicritical}, and shown in Fig~\ref{PhaseDiagram}. The phase diagram is composed of two ferromagnetic phases with magnetic moment in $x$-direction (FM$_x$) and $y$-direction (FM$_y$) and a paramagnetic phase (PM), along with multiple phase boundaries and tricritical points. 

\subsection{Quench protocol and operator pool}
We consider the linear ramp protocol: $\gamma(t) = 1 - \frac{2t}{T}$ with $0 \leq t \leq T$, as shown in Fig.~\ref{PhaseDiagram}. The longitudinal magnetic field is set to $h_z = -0.7$ and $1.6$ to avoid degenerate ground states that occur along the vertical line $h_z = 0$. In the adiabatic limit, $T \rightarrow \infty$, and in the thermodynamic limit, $N \rightarrow \infty$, the system evolves from the FM$_x$ phase, crosses a phase boundary and enters the FM$_y$ phase. In the following, we choose a finite quench speed of $T=3$, such that nontrivial spin dynamics is developed in the linear ramp process. System sizes $N\in [2, 11]$ have been considered. We restrict the operator pool to Pauli terms that appear in the Hamiltonian~\eqref{eq: XYZ}. The quantum gates to be applied on the reference state $\ket{\Psi_0}$, which are the exponentials of the scaled Pauli terms appearing in Eq.~\eqref{eq: ansatz}, include single-site and two-site Pauli rotation gates. Since the Hamiltonian~\eqref{eq: XYZ} only contains nearest-neighbor coupling terms, an expanded operator pool, which includes a complete set of two-site Pauli terms, has also been investigated. Interestingly, we find that AVQDS with the expanded pool generally produces longer variational circuits, although the simulation results are of the same accuracy. The reason can be attributed to the fact that the list of new McLachlan distances $\{ L^{2}_{\nu} \}$, see Eq.~\eqref{eq: L2min} and Fig.~\ref{algorithm}, often contains almost degenerate minimal values among several operators. The ``biased'' choice of operators in the physical pool shows some advantage due to direct connections with the Hamiltonian, which governs the quantum dynamics of the system. 

\subsection{Simulation results}
\label{avqds_results}
To characterize the time-evolution of the quantum state $\ket{\Psi(t)}$ under a quench with inverse speed $T=3$, we calculate the instantaneous total energy and spin correlation functions. Results are presented in Fig.~\ref{escgates}(a-d) for the LSM model with N=8, where the system is further evolved for an additional time period of $T$ after the linear ramp under the final Hamiltonian $\h(T)$, in order to assess the flexibility of the variational ansatz in describing the post-quench dynamics. The upper (lower) panels show results for system size $h_z = -0.7 (1.6)$. In Fig.~\ref{escgates}(e,f), we show a comparison of the number of two-qubit gates in the quantum circuits that describe the Trotterized dynamics and AVQDS. The AVQDS results, shown as solid lines, are in excellent agreement with the exact data indicated by symbols over the full time range. At the end of linear ramp $t=T$, the ansatz fidelity $f \equiv \abs{\ov{\Psi[\bth(T)]}{\Psi_\text{exact}(T)}}^2$ is beyond $99.9\%$. Numerically exact results for reference are obtained by propagating the state by matrix exponentiation, $\ket{\Psi[t+\delta t]} = e^{-i\h[t] \delta t}\ket{\Psi [t]}$, for a fine time mesh with step size $\delta t_{\text{exact}} = 5\times 10^{-4}$. In the adiabatic limit, the results for the energy and spin correlation functions are symmetric under the transformation $t \rightarrow T-t$ for $0\leq t \leq T$, but the finite quench speed breaks that symmetry. In the initial state, the dominant spin correlations are $S^x S^x$, as exemplified by the correlations between the first and second site, $S^x_0 S^x_1$, and between the first and the last site, $S^x_0, S^x_{N-1}$. During time-evolution, these correlations decrease as the parameter $\gamma$ decreases, which reduces the strength of the $S^x_i S^x_{i+1}$ interactions. Instead, the spin correlations among the $y$-components increases, as exemplified by $S^y_0 S^y_1$. The system energy increases and reaches a maximum close to $\gamma = 0$, where the phase transition occurs in the thermodynamic limit. For $\gamma < 0$, the energy begins to decrease as the $S^y_i S^y_j$-spin correlations continue to grow. Duo to the nonadiabatic finite speed quench, the long-range correlation $S^y_0, S^y_{N-1}$, which requires longer time to establish, remains far from equilibrium value at $t=T$. Although the dynamical spin correlations are remarkably distinct between the two $h_z$ points of parameter space, the adaptive variational circuits reach about the same complexity of 50 two-qubit Pauli rotation gates at the end of linear ramp. The post-ramp dynamics in the following time period of $T$ slightly expands the ansatz by $3$ two-qubit gates.

To compare the circuit complexity of AVQDS with the Trotter approach, we perform Trotterized simulations with a uniform time step $\delta t = 5\times 10^{-3}$, applying a series of one and two-qubit Pauli rotation gates to the state. To characterize the accuracy of the dynamics simulations, we evaluate the standard deviation of an observable $\hat{O}$ along the linear ramp dynamical path for $0\leq t \leq T$ according to
\be
s = \sqrt{\frac{1}{N_t - 1} \sum_{t} \left(\av{\hat{O}}_t^\text{Trotter} - \av{\hat{O}}_t^\text{exact}\right)^2},
\ee
where the summation is over the entire time mesh of dimension $N_t$ for Trotter simulations. Note that the expectation value $\av{\hat{O}}_t$ at a specific time $t$ in the simulation is calculated rigorously without any noise, which corresponds to infinite repeated measurements of the observable $\hat{O}$ on ideal fault-tolerant quantum computers. The standard deviation is $0.003$ for both energy and the shown spin correlation functions for a system size of $N=8$. As discussed previously in section~\ref{dtheta_max}, the maximum allowed parameter step size $\delta \theta_\text{max}$ is the proper controlling factor in AVQDS, replacing $\delta t$ that is the relevant quantity in Trotter dynamics simulations. To properly benchmark AVQDS and obtain a reasonable comparison of the resulting circuit complexities, we thus set $\delta \theta_\text{max} = 5\times 10^{-3}$ equal to the Trotter timestep $\delta t$ in the AVQDS simulations. This results in a standard deviation of $0.003$ for spin correlation observables, and $0.010$ for the energy, which is nevertheless about three times bigger than the Trotter energy error.

The number of two-qubit gates is the defining factor for practical quantum computing in the NISQ era. As shown in Fig.~\ref{escgates}(e) and (f), we compare the number of two-qubit gates contained in the variational ansatz and in the Trotterized circuits as a function of time $t$. Importantly, the AVQDS simulations require up to two orders of magnitude fewer two-qubit gates than Trotter simulations. In contrast to the linear growth of Trotter circuits, the AVQDS circuits mainly grow in the initial quench stage, and approach a plateau as the system evolves further. For $N=8$, we find a plateau value of about $100$ CNOTs is sufficient to follow the spin dynamics to the final simulation time, whereas Trotterized circuits require execution of about $10^4$ CNOTs. This implies that the variational circuit has gained sufficient expressibility to represent the relevant manifold of the Hilbert space for the linear-ramp dynamics studied here.

\begin{figure}[t]
	\centering
	\includegraphics[width=0.9\columnwidth]{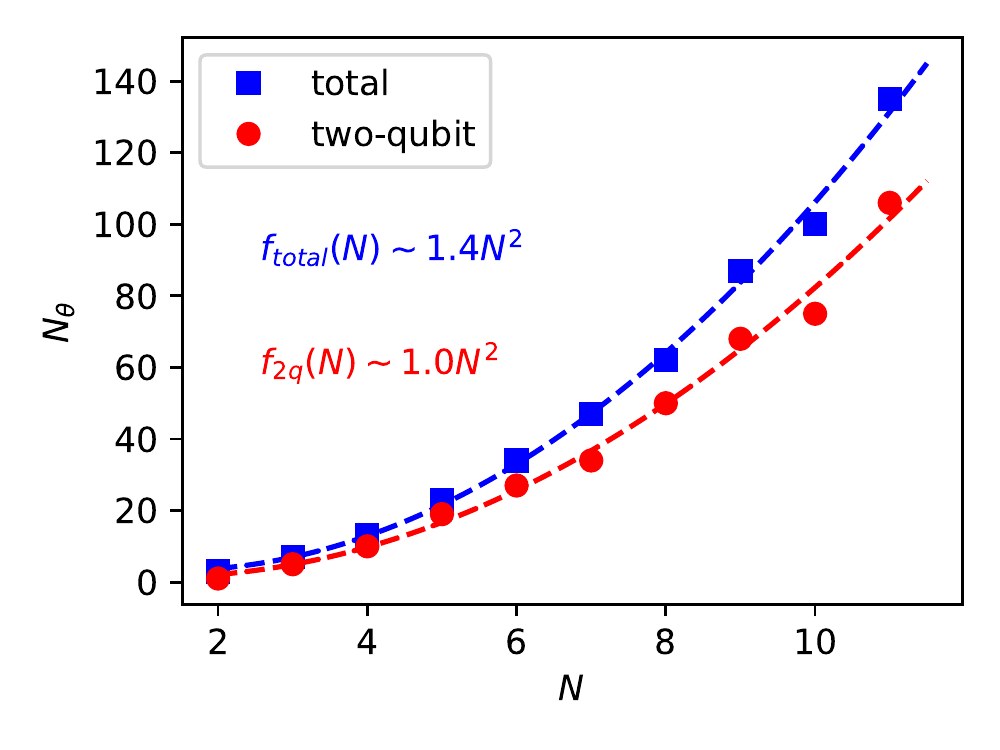}
	\caption{
	\textbf{Saturated number of variational parameters ($N_{\bth}$) as a function of LSM model size $N$.} The total number of variational parameters for AVQDS simulations after saturation is reached (blue squares) grows quadratically with the number of sites $N$. The corresponding number of variational parameters associated with two-qubit Pauli rotation gates (red dots) also shows quadratic scaling behavior $\propto N^2$, yet with a smaller prefactor. 
	}
	\label{nvars}
\end{figure}

\begin{figure*}[t]
	\centering
	\includegraphics[width=\textwidth]{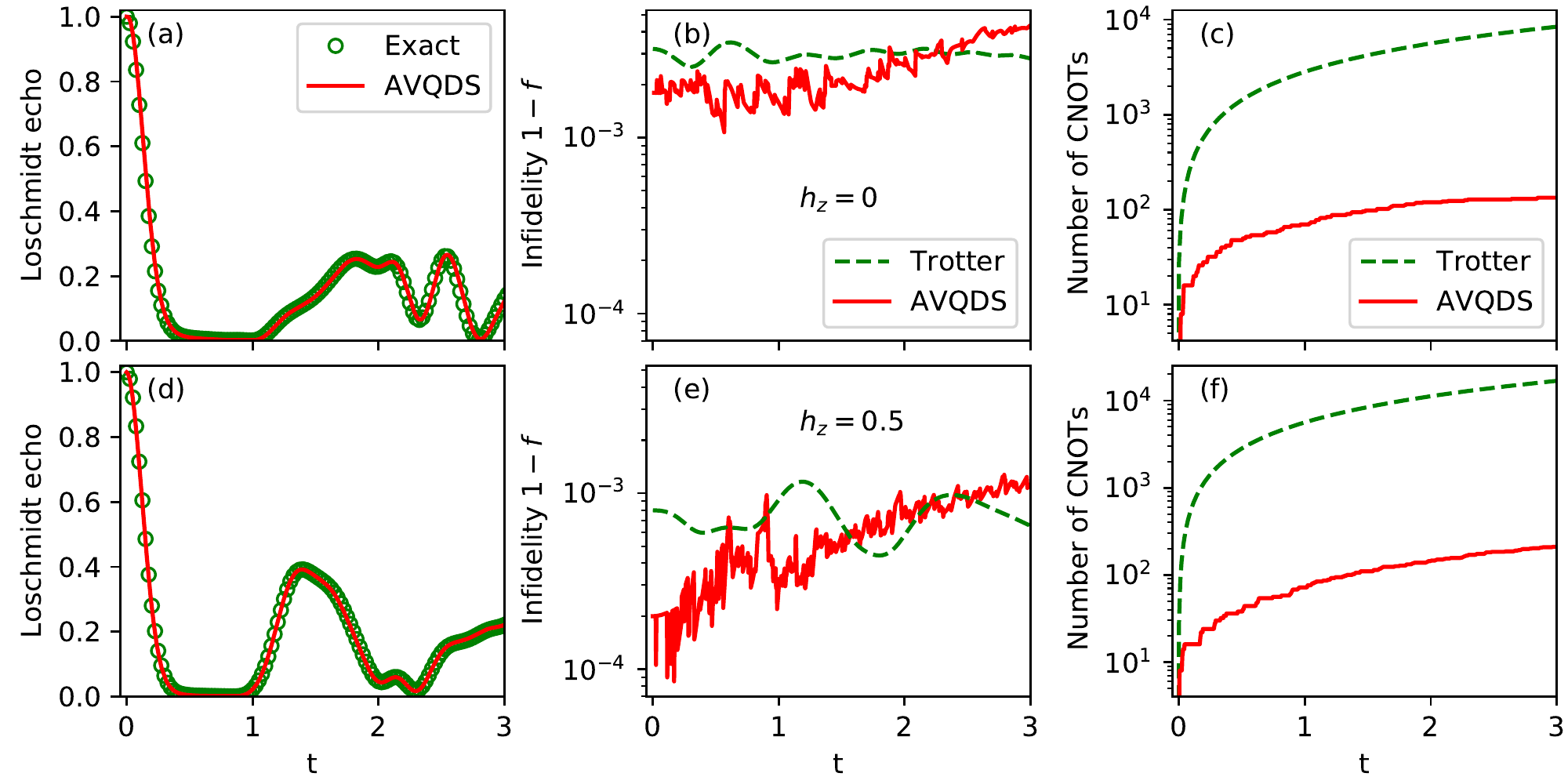}
	\caption{
	\textbf{Quench dynamics of eight-site mixed-field Ising model with periodic boundary conditions.} The time-dependent Loschmidt echo, infidelity $1-f \equiv 1-\abs{\ov{\Psi[\bth(t)]}{\Psi_\text{exact}(t)}}^2$, and number of CNOTs $N_\text{cx}$ are plotted in upper panels for $h_z=0$ (integrable), and lower panels for $h_z=0.5$ (nonintegrable). The Loschmidt echoes (return probabilities) in (a) and (d) from AVQDS show excellent agreement with exact calculations. The infidelity plots in (b) and (e) further quantify the AVQDS accuracy with a fidelity $\geq 99.5\%$. As shown in (c) and (f), Trotter circuits are about two orders of magnitude deeper than AVQDS circuits for similar simulation accuracies, shown in (b) and (e). At $t=3$, $N_\text{cx}$ reaches 134 and 210 for the integrable and nonintegrable models, respectively.
	}
	\label{fig: quench}
\end{figure*}

\subsection{System-size scaling of circuit complexity}
\label{sec: scaling}
To reach the goal of performing scalable quantum dynamics simulations on quantum devices, it is crucial to estimate how the required quantum resources scale with the system size $N$. Although this scaling depends on the complexity of the dynamical problem studied and is therefore model dependent, it is instructive to investigate the scaling of the required resources with system size $N$ for the LSM model. As shown in Fig.~\ref{nvars}, the final number of parameters of the variational ansatz in AVQDS simulations scale with $N^2$. Specifically, we find that the total number of variational parameters, which are associated with single and two-qubit Pauli rotation gates, scales as $1.16\, N^2$. The number of variational parameters that are associated with two-qubit Pauli rotation gates scales as $0.78\, N^2$. 

The AVQDS approach utilizes an inherent Trotter-type structure for the variational ans\"atze with unitaries constructed based on Hamiltonian Pauli terms. It is conceptually different from the random parameterized quantum circuit optimization approach, where the cost function gradients can become exponentially small with increasing number of qubits~\cite{mcclean2018barren, grant2019initialization, arrasmith2020effect}. In the context of adiabatic state preparation, AVQDS can lead the system to the ground state without resorting to explicit high-dimensional optimization or cost function gradients. Therefore, barren plateaus of cost functions associated with random variational circuits are unlikely to constitute a problem for AVQDS simulations. As further numerical evidence, the element-wise maximal absolute value of the vector $V$ defined in Eq.~\eqref{eq: V}, which is closely related to the cost function gradient, remains close to $0.28$ in AVQDS simulations of the LSM model as the number of sites is increased from $4$ to $8$.

The above AVQDS calculations start with the ground state $\Psi_0$ of the Hamiltonian $\h_0$ at $t=0$ in the FM$_x$ phase. Due to the finite transverse field $h_z$, this state is not a tensor-product state and preparation of the initial state is therefore non-trivial. While it is convenient to initialize any state vector in classical simulations, it is not straightforward to prepare an entangled state on quantum computers. As discussed previously in Section~\ref{adapt-vqe}, we here adopt the qubit-ADAPT-VQE method~\cite{vqe_qubit_adaptive} for the initial state preparation. We use a reference product state with all spins aligned in $\sigma_z=1$ and an expanded operator pool that includes all one- and two-qubit Pauli terms. For system size $N=6$, we find that an ansatz with 20 variational parameters,  which are associated with two-qubit Pauli rotation gates, is able to variationally represents $\ket{\Psi_0}$ with unit overlap up to the $7^\text{th}$ decimal place. At the end of the AVQDS simulation, the number of variational parameters associated with two-qubit Pauli rotation gates has increased to $45$, which is slightly smaller than a rough estimation of $20 + 0.78 N^2 \approx 20+28$ for $N=6$. Here $20$ is the number of variational parameters in the initial qubit-ADAPT-VQE ansatz as mentioned above, and $28$ is the number of variational parameters added during time-evolution. 
\label{adapt-vqe-avqds}

\section{Sudden quench dynamics of the mixed-field Ising model}
\label{sec: mfim}
To further benchmark the AVQDS approach for nonintegrable dynamics, we perform sudden quench dynamics simulations in the mixed-field Ising model (MFIM):
\be
   \h = -J\sum^{N-1}_{i=0} \hat{Z}_i \hat{Z}_{i+1} + \sum^{N-1}_{i=0} \left( h_x\, \hat{X}_i + h_z\, \hat{Z}_i\right),
\ee
with $\hat{Z}_N=\hat{Z}_0$ for periodic boundary conditions. In the following, we measure energy in units of $J=1$. This model is integrable for $h_z = 0$, where it becomes the transverse-field Ising model (TFIM), but is nonintegrable when both $h_z$ and $h_x$ are finite. Initially, the system is prepared in the ordered state $\ket{\Psi_0}=\ket{\up \dots \up}$, which is a ground state of the MFIM in the absence of magnetic fields $h_x = h_z = 0$. We consider two sudden quench protocols: (A) quenching to the TFIM with $h_x = -2.0$ at $t=0$, and (B) quenching to the MFIM with $h_x = -2.0, h_z = 0.5$ at $t=0$. Protocol A has been used in the study of dynamical quantum phase transitions~\cite{heyl2013dynamical, heyl2018dynamical}. Protocol B allows us to compare the performance of AVQDS for integrable and nonintegrable models.

The Loschmidt echo, defined as the probability of a time-evolving system to return to its initial state, is a central concept in the study of dynamical quantum phase transitions~\cite{heyl2013dynamical, heyl2018dynamical}. The simulations presented here starts with the initial state $\ket{\Psi_0}=\ket{\up \dots \up}$, and the Loschmidt echo can be written as $\Lag(t) = \abs{\Mel{\Psi_0}{e^{-i\h_f t}}{\Psi_0}}^2$. 
The time-dependent Loschmidt echo calculated from AVQDS is plotted in Fig.~\ref{fig: quench} for (a) the integrable TFIM and (d) the nonintegrable MFIM. Both cases are in excellent agreement with exact results. To better quantify the simulation accuracy, the infidelity $1-f\equiv 1-\abs{\ov{\Psi[\bth(t)]}{\Psi_\text{exact}(t)}}^2$ is shown in panels (b) and (e), indicating that the fidelity $f$ is generally higher than $99.5\%$. To make a comparison with naive Trotter dynamics, we performed Trotter simulations with a time step size chosen to provide fidelities comparable to those obtained with AVQDS [see green dashed lines in (b) and (e)]. In panels (c) and (f), we plot the number of CNOTs $N_\text{cx}$ used in the Trotter and variational AVQDS circuits. Consistent with the results of Sec.~\ref{sec: results}, we find that Trotter circuits have about two orders of magnitude more two-qubit gates than the AVQDS circuits. A tendency towards circuit depth saturation becomes noticeable for the integrable TFIM ($h_z=0$) for simulation times $t\gtrsim 2$. The parameterized circuit for $\ket{\Psi[\bth(t)]}$ reaches $134$ CNOTs at final simulation time $t=3$, slightly larger the $N=8$ LSM simulation result. Moving from the integrable TFIM to the nonintegrable MFIM by introducing finite field $h_z$, the AVQDS circuit increases $N_\text{cx}$ modestly from $134$ to $210$ at $t=3$.

\begin{figure*}[ht]
	\centering
	\includegraphics[width=\textwidth]{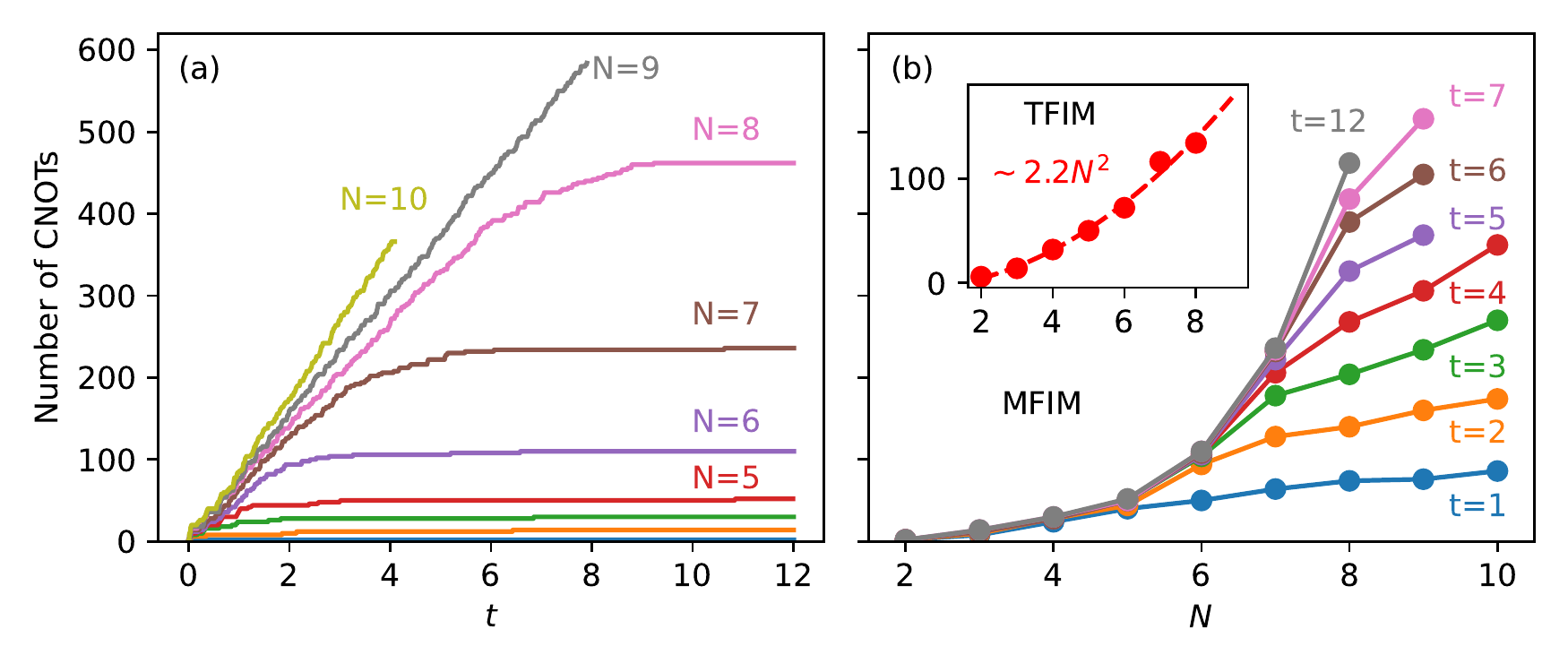}
	\caption{
\textbf{Number of CNOTs $N_\text{cx}$ in the AVQDS circuits for the MFIM as a function of simulation time $t$ and system size $N$.} (a) The AVQDS circuit depth for MFIM initially grows linearly with $t$, followed by a slowdown toward saturation. Times up to $t=12$ are considered for $N\leq 8$, with shorter final times shown for $N=9, 10$. Clearly for $N \le 8$, the circuit begins to saturate at a critical time $t_\text{s} < 12$. (b) Vertical cuts of panel (a) at constant times for $1\leq t\leq 7$ and $t=12$. Inset: Quadratic scaling of the saturated $N_\text{cx}$ for AVQDS simulations of the integrable TFIM.
	}
	\label{fig: ngates2}
\end{figure*}

In Fig.~\ref{fig: ngates2}, we consider the scaling of the number of CNOT gates $N_\text{cx}$ in the AVQDS circuit with time $t$ (panel a) and system size $N$ (panel b). This is the number of CNOT gates required to build the post-quench state $\ket{\Psi[\bth(t)]}$ in the MFIM. Importantly, we observe an initially linear growth, $N_{\text{cx}} \propto t$, that crosses over into saturation at a system size dependent timescale $t_\text{s}(N)$. The initial growth resembles the behavior of the entanglement entropy in the post-quench regime~\cite{kim2013ballistic}. Since the number of CNOTs in the circuit is proportional to the number of variational parameters, this (polynomial) growth is notably slower than the exponential growth of the number of parameters needed in simulations using matrix-product states (MPSs). This is a direct manifestation of the \emph{complexity window}, which describes the phenomenon that quantum circuits can generate states with high entanglement more efficiently, i.e. with less parameters, than MPSs~\cite{brandao2019models, lin2021real}. More precisely, states in the complexity window are highly entangled yet can be represented by a quantum circuit that grows only polynomially in time~\cite{poulin2011quantum}. Figure~\ref{fig: ngates2} thus shows that such efficient circuits can be automatically generated within AVQDS. 

Specifically, Fig.~\ref{fig: ngates2}(a) shows $N_\text{cx}$ of the AVQDS circuits for the MFIM at system sizes $2\le N \le 10$ with simulation times up to $t=12$, except for $N=9$ and $10$, which use shorter final times. We observe that $N_\text{cx}$ for $N \le 8$ saturates within $t=12$. The critical time $t_\text{s}$ for saturation increases with $N$. Panel (b) depicts $N_{\text{cx}}$ as a function of system size $N$ for different fixed times $t$ (vertical cuts of the data in panel (a)). While the saturated circuit depth, as measured by $N_\text{cx}$ at $t=12$, appears to scale beyond quadratically (with $N$) for $2\le N \le 8$ due to the increased complexity of the nonintegrable model, the exact order of the scaling cannot be obtained due to the small size of the data set. Similar quality of fitting is obtained using a power-law function $a N^{\alpha}$ with $\alpha \approx 4.8$ and an exponential function $b(e^{N/\beta}-1)$ with $\beta \approx 1.4$. Let us now consider the practically relevant question of how the circuit depth grows with $N$ for a sequence of fixed simulation times $1 \le t\le 7$. Because the saturation time $t_\text{s}$ increases rapidly with $N$, the equal-time cut generally exhibits a crossover behavior of $N_\text{cx}$ from the initial super-quadratic growth with small $N$ to approximately linear growth with large $N > N_\text{c}(t)$ as the circuits cross over into the presaturation regime (where $t < t_s(N)$). The crossover size $N_\text{c}(t)$ grows slowly with $t$. This suggests that practical dynamics simulations of generic quantum models out to fixed final times within the AVQDS approach remains scalable with increasing $N$. Finally, in the inset of Fig.~\ref{fig: ngates2}(b), we show that  $N_\text{cx}$ after the saturation time $t_\text{s}$ in the integrable TFIM grows approximately as $N^2$, which is slower than the $N^{4.8}$ observed for the MFIM case. The quadratic scaling is similar to the results for the LSM chain shown in Fig.~\ref{nvars}, suggesting that it is related to the integrability of the TFIM and the LSM models. 



\section{Conclusion}
\label{sec: conclusion}
We present a novel adaptive variational approach, AVQDS, to perform quantum dynamics simulations in many-body fermionic and spin models. It builds upon the theory of variational quantum dynamics according to McLachlan's variational principle. The key novelty of the presented AVQDS method is to adaptively and dynamically expand the variational ansatz during time evolution. This allows to account for the changing nature of the time-evolved wavefunction and results in highly accurate variational quantum dynamics simulations. Expansion of the ansatz is controlled by setting a threshold of the maximum allowed McLachlan distance $L^2$. This distance describes the difference after one time step between the exact time-evolution of the variational state, as described by the von-Neumann equation, and the time evolution obtained from classically propagating the variational parameters in time. The AVQDS approach does not involve complex optimization in a high-dimensional parameter space. 

To benchmark and study the performance of the approach, we use AVQDS to simulate spin dynamics in the integrable LSM model under a finite-rate quantum quench, and in the nonintegrable MFIM under sudden parameter quenches. We consider system sizes up to $N=11$ sites. The AVQDS simulations are shown to be in excellent agreement with numerically exact results for local observables, total energy, and wavefunction overlaps. The depth of the resulting AVQDS variational circuits, as characterized by the number of two-qubit CNOT gates $N_{\text{cx}}$, is shown to saturate to values smaller than Trotterized circuits by up to two orders of magnitude for $N=8$. We find that $N_{\text{cx}}$ after saturation at the end of the linear ramp scales as $N_{\text{cx}} \propto N^2$ for both the LSM and the TFIM model, suggesting that this polynomial scaling is linked to the integrability of models. For quench dynamics simulations of the nonintegrable MFIM, the saturated number of two-qubit gates in the AVQDS circuits scales as a higher-order polynomial $N_{\text{cx}} \propto N^5$ approximately, as expected from the increased complexity of the model. For fixed times $t < t_s(N)$, however, we observe that the number of two-qubit gates in the AVQDS circuits reduces to approximately linear growth $N_{\text{cx}} \propto N$ as system size increases, implying the scalability of AVQDS simulations in practice. Finally, we find that the complexity of the AVQDS circuits at fixed $N$ scales initially linearly with time $N_{\text{cx}} \propto t$, showing that these circuits can efficiently capture the rapid growth of entanglement under nonintegrable dynamics in the system. 

We envision that the AVQDS approach will have wide applications in the growing field of quantum dynamics and far-from-equilibrium physics. In addition to directly simulating dynamics in other spin and fermionic models, AVQDS can be used as an impurity dynamics solver for quantum embedding approaches for dynamics simulations of large and infinite lattice models~\cite{schiro10tdga, RMPNonequilibriumDMFT, kretchmer18tddmet, gqce}. An open question to further explore the finite-size and finite-time scaling of the AVQDS circuit depth and relating it to the entanglement content of the time-evolved state. The prospect of adaptively and automatically generating polynomial depth circuits that generate highly entangled states is intriguing and warrants further investigation. 
Another important future research direction is to study the noise resilience of the algorithm and noise mitigation strategies, in particular when implementing AVQDS on NISQ QPUs. For the preparation of the initial state of the dynamics simulation, we have explicitly shown that AVQDS can be easily combined with the known qubit-ADAPT-VQE method. Finally, AVQDS can be generalized from real time to imaginary time axis~\cite{AVQITE}, which offers a novel efficient approach to finding ground states of Hamiltonian systems, or to a wider range of optimization problems of static cost functions in the field of machine learning.

\section*{Acknowledgements}
We are grateful for discussions with Anirban Mukherjee. This work was supported by the U.S. Department of Energy (DOE), Office of Science, Basic Energy Sciences, Materials Science and Engineering Division. The research was performed at the Ames Laboratory, which is operated for the U.S. DOE by Iowa State University under Contract No. DE-AC02-07CH11358.

\bibliography{refabbrev, ref}

\end{document}